# Distributed state estimation: a novel stopping criterion

Sajjad Asefi*. Sergei Parsegov** Elena Gryazina***

*Center for Energy Science & Technology, Skolkovo Institute of Science & Technology, Moscow, Russia
*(Sajjad.Asefi@skoltech.ru) **(S.Parsegov@skoltech.ru) ***(E.Gryazina@skoltech.ru)*

Abstract: Power System State Estimation (PSSE) has been a research area of interest for power engineers for a long period of time. Due to the intermittent nature of renewable energy sources, which are applied in the power network more than before, the importance of state estimation has been increased as well. Centralized state estimation due to the complexity of new networks and growing size of power network will face problems such as communication bottleneck in real-time analyzing of the system or reliability issues. Distributed state estimation is a solution for the mentioned issues. There are different implementation methods introduced for it. The results of the paper are twofold. First, we examined different approaches to distributed PSSE (DPSSE) problem, according to most important factors like iteration number, convergence rate, data needed to be transferred to/from each area and so on. Next, we proposed and discussed a new distributed stopping criterion for the methods and above-mentioned factors are obtained as well. Finally, a comparison between the total efficiency of all applied methods is done.

## 1. INTRODUCTION

State estimation (SE) is an essential part of the current power system. The major benefits of state estimation are estimation of network parameters based on redundancy in measurements, bad data and topology errors detection, estimating meter measurements for missing or delayed data (Wood et al. 2013). Advance state estimation can improve monitoring and control of the power system in the case of contingency occurring. Providing reliable and complete information is the main duty of the state estimator, which has a great importance for online operations and control systems that guarantee security of the power grid (Gomez-Exposito et al. 2018). In other words, the main function of the state estimator is to identify the system state by minimizing a specific criterion based on recent system measurements (Ahmad et al. 2018). In addition, the increasing presence of renewable energy sources (RESs) in the power grid requires state estimation to be more precise and fast due to RESs intermittent nature.

Power system state estimation problem has been a matter of concern back in 1970s, when first Schweppe et al. (1970 a, b, c) developed a model to solve this problem. Traditionally PSSE is done in a centralized manner, in which a single unit collects all data ('data' here refers to measurement unit data and system parameters such as line impedance) of the system, and an optimization technique is used to solve the PSSE problem. Later on, not only for state estimation but also all optimization problems, distributed solutions attracted researchers interest. There have been plenty of investigations about DPSSE in literature.

In (Conejo et al. 2007), a simple yet general multi-area state estimation method is described, which can provide a correct estimation of the system states. While the system is in connection with other systems, they only interchange a small amount of border data without any need for data to be processed or manipulated. Kekatos et al. (2013 a, b) have provided a novel algorithm for DPSSE using alternating direction method of multipliers (ADMM). As stated by the authors, besides applying conventional least squares state estimation, the proposed distributed process constructs a robust state estimator. In (Minot and Li 2015) and (Minot et al. 2016) a DPSSE using matrix splitting method is presented. First, in (Minot and Li 2015) authors evaluated the method for DC state estimation, but later in (Minot et al. 2016) they provided a distributed Gauss-Newton method for AC state estimation. It is to be noted that both papers have considered PMU (phasor measurement units) measurements in addition to conventional SCADA (supervisory control and data acquisition) data. All the mentioned approaches are different in terms of applicability.

Evidently, any optimization procedure needs certain stop-ping criteria. Usually this is done considering an accuracy limit of objective function, step tolerance and optimality tolerance. Due to the increase in measurement units and computational demand for growing power networks, there has been a need to apply DPSSE to speed up the processing time by spreading the computational tasks among different control units. Additionally, decentralization decreases the possibility of communication bottleneck, which is more probable to happen for a huge system centralized state estimation (Kekatos et al. 2017). In some cases, due to data privacy and cybersecurity reasons, the state estimation must be distributed, e.g., when areas ('area' here refers to a partition of the power system) are different countries. In contrast to centralized PSSE, novel distributed stopping criterion should be proposed and applied, since we have local solvers associated with each area for distributed PSSE because whole information is not available to make a decision. Moreover, a proper choice of distributed stopping criterion may also reduce the amount information sent through the communication channel.

In this paper, we examined recent and well-known methods mentioned above by IEEE standard test systems, and the performance of these methods from different viewpoints, such as data needed to be transferred, computation time, error of obtained solution compared to centralized one and convergence rate, are compared. This comparison serves for the main goal of the paper which is development and implementation of stopping criterion for each area, which decreases communication burden and needed data to be transferred to/from each area and consequently, decreases iteration number for getting a reasonable solution. To best of the authors' knowledge, the effect of stopping criterion on DPSSE has not been addressed in the literature. Finally, we studied efficiency of proposed approach using different distributed optimization methods.

The paper is organized as follows. Section II discusses about the main problem formulation and objective function. Additionally, well-known DPSSE methods are presented in this section. In Section III the algorithm related to consideration of stopping criterion is discussed. Section IV presents the simulation results and discussions. And finally, in Section V the paper is concluded.

## 2. DPSSE METHODS AND FORMULATION

The type of PSSE we have considered in this work is Static SE. Static SE function is used to monitor the system during normal operation. Given the network model, all measurement types (such as power flows, power injections, virtual measurements (which is based on physical assumption of the system, like zero net power injection at buses with no load or generation) and pseudo-measurements, i.e. measurement values that are predicted based on historical data) can be expressed as a function of the system states. Formulation of the SE problem is based on the concept of maximum likelihood estimation (Gomez-Exposito et al. 2018). One can write the equation in matrix format, which has been provided below:

$$\min J(x) = \min[z - h(x)]^T \times R^{-1} \times [z - h(x)], \quad (1)$$

where $R$ is a covariance matrix, $z$ is the measurements matrix and $h(x)$ is a function that shows the relation between measurements and state variables. In order to solve this optimization problem, the solution must pursue first order optimality condition. There are two points of view on the problem. One is DC SE, in which the $h(x)$ is a linear function (that means measurements have linear relation with state variables) and we can do the matrix calculation directly. The other one is AC SE, in which $h(x)$ is nonlinear and we must use other methods like Newton or Gauss-Newton, to solve it. This paper deals with only DC SE. As stated before, if we consider a linear relation between measurement units and state variables ($h(x) = H \times x$, where $H$ indicates the measurement matrix and $x$ represents all voltage phase angles ($\theta$) of the power system buses), the solution to (1) can be obtained in a single step as follows, meeting the first order optimality condition:

$$x^* = (H^T R^{-1} H)^{-1} \times H^T R^{-1} z. \quad (2)$$

In order to calculate (2) we need to access all data in the system by a single (or centralized) control unit. As mentioned before, issues like increase in the number of measurement units in power system, communication bottlenecks and data security, are the main reasons which leads power system to utilize distributed approaches. Further, a brief overview about recent and well-known DPSSE known approaches are provided.

### 2.1 Matrix splitting

Based on (Minot and Li 2015), In order to solve problem (1) in a distributed manner, one can use matrix splitting method and after a certain number of iterations the answer converges to the centralized solution. The main equation of matrix splitting for a problem of $Ax = y$ is:

$$x^{t+1} = M^{-1} N x^t + M^{-1} y. \quad (3)$$

$A$ is written as the sum of an invertible (or diagonal) matrix $M$, and a matrix $N$; i.e. $A = D + E$, or $A = M + N$ so that $M = D + E'_{ii}$ and $N = E - E'_{ii}$. Note that $D$ contains diagonal arrays and $E$ contains off-diagonal arrays of matrix $A$. And $E'_{ii}$ is a diagonal matrix which is defined as follows:

$$E'_{ii} = \alpha \times \sum_{j=1}^{n} |E_{ij}|, \quad (4)$$

that we have assumed $\alpha = 1$ for simplicity. It is to be noted that (3) converges if the spectral radius of $M^{-1}N$ matrix be less than 1 ($\rho(M^{-1}N) < 1$). Using (3) iteratively results in convergence to the system's ($Ax = y$) final solution ($x^*$).

### 2.2 Gossip based method

In this section, we are going to introduce another matrix splitting approach, which is discussed in (Frasca et al. 2015). The authors considered data transfer between areas in asynchronous manner (so called gossip communication protocol), and the results show that many iterations are needed to get a reasonable convergence. Now, let us briefly describe how the method works.

In DC model, the state estimation problem in the least squares setting can be formulated as (1) and (2). Based on what has been stated in (2), this problem has a closed-form solution. Let's assume, $L = H^T R^{-1} H$ and $u = H^T R^{-1} z$. One way to compute this solution $x^*$ is through the gradient based iterative algorithm:

$$x(k+1) = (I - \tau L)x(k) + \tau u. \quad (5)$$

The parameter $\tau$ is selected from the interval $(0, 2\|L\|^{-1})$; such a $\tau$ guarantees the matrix $I - \tau L$ to be Schur stable (i.e. the iterative method converges). At each iteration, a set of pair areas, randomly (based on uniform probability distribution) will be selected to update the common variables.

## 2.3 Decomposition method

In this section, we are going to discuss the method provided in (Conejo et al. 2007). This method that can be categorized as a decomposition method, applies simple power flow and power injection equations. Based on what have been stated in literature, it is clear that the single-area (general) state estimation problem and multi-area state estimation problem can be formulated as follows for DC state estimation:

$$\min_{x_a} J_a(x_a) + \sum_{b \in \Omega_a} J_{ab}(x_a, \tilde{x}_b), \quad (6\text{-a})$$

$$J_a(x_a) = \sum_{i \in \Omega_a^P} \omega_{a,i}^P (P_{a,i}^m - P_{a,i}(.))^2 \\ + \sum_{(i,j) \in \Omega_a^{PF}} \omega_{a,ij}^{PF} (P_{a,ij}^m - P_{a,ij}(.))^2, \quad (6\text{-b})$$

$$J_{ab}(x_a, \tilde{x}_b) = \sum_{i \in \Omega_{ab}^P} \omega_{ab,i}^P \left(P_{ab,i}^m - P_{ab,i}(.)\right)^2 \\ + \sum_{(i,j) \in \Omega_{ab}^{PF}} \omega_{ab,ij}^{PF} (P_{ab,ij}^m - P_{ab,ij}(.))^2 \\ + \sum_{i \in \Omega_{ab}} \omega_{b,i}^x (\tilde{x}_{b,i} - x_{b,i})^2, \quad (6\text{-c})$$

where $J_a(.)$ is weighted measurement error function for area $a$ involving only state variables of area $a$; $J_{ab}$ is weighted measurement error function for area $a$ involving state variables of area $a$ and $b$; and $\Omega_a$ is the set containing indices for all neighbouring areas of area $a$; $\omega$ is weighting factor; $P_{(.),i}^{(.)}$ is active power injection at bus $i$; $P_{(.),ij}^{(.)}$ is power flow in between bus $i$ and $j$; and $m$ indicates measurement. To solve (6), we applied *MATLAB* built-in solver (Sequential quadratic programming (SQP) via *MATLAB R2018b* on a computer with Intel(R) Core i5 processor and 8 GB of RAM) and the results are provided in this paper.

## 2.4 ADMM

In (Kekatos and Giannakis 2013) a new method was developed for solving distributed PSSE, which is based on ADMM presented in (Boyd and Vandenberghe 2004). ADMM increases existing PSSE solvers performance, and convergence of the method to its centralized counterpart is guaranteed, even if there is lack of local observability. Although, ADMM can be considered mathematically as a decomposition method, but due to recently increased application of this method for DPSSE, we decided to consider it separately as well. In general, the DPSSE problem can be formulated as:

$$\min_{x_k} \sum_{k=1}^{K} f_k(x_k), \quad (7\text{-a})$$

$$x_k[l] = x_l[k], \forall l \in N_k, \forall k, \quad (7\text{-b})$$

where $N_k$ is the set of areas sharing states with area $k$ and $x_{k,l}$ is auxiliary variable introduced per pair of interacting areas $k$, $l$.

The constraint forces neighboring areas to consent on their shared variables. Augmented Lagrangian function is as follows:

$$L(\{x_k\}, \{x_{kl}\}; \{v_{kl}\}) \\ := \sum_{k=1}^{K} [f_k(x_k) \\ + \sum_{l \in N_k} (v_{k,l}^T (x_{k[l]} - x_{kl}) + \frac{c}{2} \|x_{k[l]} - x_{kl}\|_2^2), \quad (8)$$

where $v_{k,l}$ is Lagrangian multiplier and $c > 0$.

$$\{x_k^{t+1}\} := \arg \min L(\{x_k\}, \{x_{kl}^t\}; \{v_{kl}^t\}), \quad (9\text{-a})$$

$$\{x_{kl}^{t+1}\} := \arg \min L(\{x_k^{t+1}\}, \{x_{kl}\}; \{v_{kl}^t\}), \quad (9\text{-b})$$

$$v_{kl}^{t+1} := v_{kl}^t + c \left(x_{k[l]}^{t+1} - x_{kl}^{t+1}\right), \forall k. \quad (9\text{-c})$$

Let's consider a system which is divided into $k$ areas. Each area collects $z_k$ measurements

$$z_k = H_k x_k + e_k, \quad (10)$$

so that $x_k$ contains system states and $e_k$ is random noise vector.

## 3. STOPPING CRITERION

The main aim of this paper is to propose and validate a distributed stopping criterion for proper DPSSE termination. In contrast to the centralized case where the whole state vector is observed and a decision for terminating objective function evaluation can be done, the distributed case has no such information. Therefore, the problem becomes crucial since may lead to the endless expensive computations and unnecessary information exchange between local computational units in areas. For practical needs it is important to propose simple yet efficient local condition for each area that once satisfied, stops evaluation within this area, fixes reached estimate and informs its neighbours about it. Therefore, the natural question that to the best of authors' knowledge has not been addressed in the literature is to consider the effect of a distributed stopping criterion. The implementation and effectiveness of distributed stopping criterion is inextricably linked with applied distributed optimization methods. A detailed performance indices are provided in tables, where the DPSSE algorithms are compared. A clear and easily implementable stopping algorithm is needed to effectively increase the overall efficiency in terms of information needed to be transferred, computational time, objective value, etc.

One of the trivial ways to stop an algorithm is to set specific number of iterations and hand out the solution when the iterations finish. Obviously, this way can not give a satisfactory result to most problems, specially state estimation which plays a vital role in power system management. In addition, the problem is not centralized anymore, which results in the fact that we need to develop and implement a simple yet effective distributed method to deal with it. The following algorithm shows the general approach to DPSSE with proposed stopping criterion:

---

**Algorithm 1** DPSSE with stopping criterion

---

Initialize the parameters for each particular method
Define area number (AN), state number (SN), each area's measurements and needed data
Specify stopping criterion parameter ($\epsilon$)
Solve the first iteration
Transmission of the needed data between each area
**While** $||x^t - x^{t-1}|| > \epsilon$ **do**
   Local computation at each area and then start sharing the common data consider the following
   **For** $k = 1$ to AN **do**
     **For** $i = 1$ to SN **do**
       **If** $||x_{i,k}^t - x_{i,k}^{t-1}|| < \epsilon$ **Then**
         $x_{i,k}$ in the next steps will be the last value and no need to transfer this data anymore
       **Else**
         keep on sending the needed data
       **end if**
     **end for**
   **end for**
**end while**

---

It is to be noted that the algorithm just shows the general concept and the practical realization of algorithms may slightly differ due to specific properties of each method.

In the following section, results related to the above mentioned question has been shown, for some well-known test cases (4, 14 and 118 bus system) using methods introduced in previous sections.

## 4. RESULTS AND DISCUSSIONS

In this section, we apply all methods to 4, 14 and 118 bus systems, and compare the results. For better understanding of the effect of stopping criterion, the graphical results have been provided as well. However, for the sake of brevity, only figures for IEEE 14 bus system (which is the most analysed system in the literature) and ADMM method (that has better results compared to other methods) is presented. Fig. 1 shows a comparison of voltage phase angles between centralized, without considering distributed stopping criterion (D-WOSC) and with considering distributed stopping criterion (D-WSC) Fig. 2 (without stopping criterion) and Fig. 3 (with stopping criterion) present the convergence curve for vector of variables (**x** is state vector and **t** indicates iteration number). Tables 1 to 3 summarize the numerical results related to the number of iterations (Iter) of different methods, error values compared to centralized solution, computational burden (CB), overall elapsed time (OT) and finally the objective function value (OV). Convergence limit $\epsilon$ was set to $10^{-6}$ for all cases. Two different scales were applied for measuring the error of each method's solution compared to the answer obtained using the centralized method.

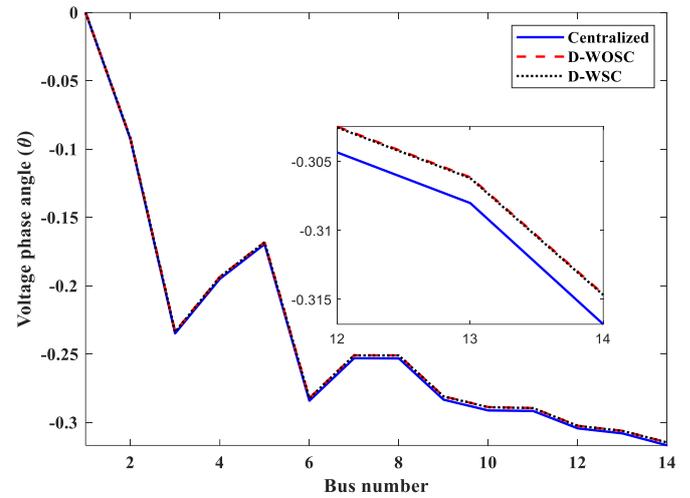

Fig. 1. Comparison of centralized and decentralized voltage phase angle (in radian) using ADMM for IEEE 14 bus system

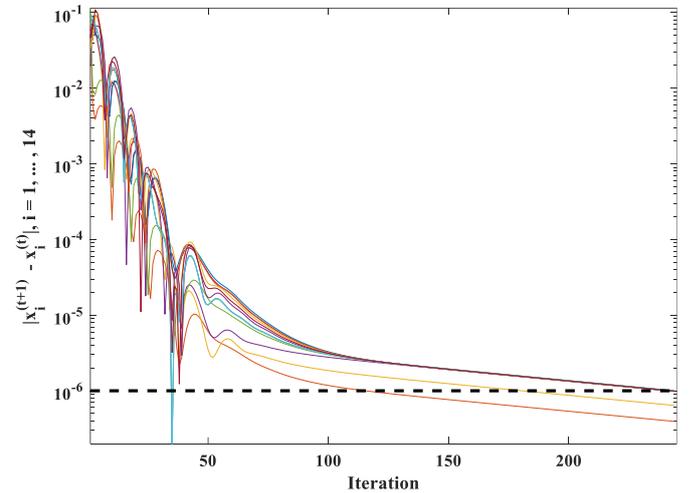

Fig. 2. Convergence curve for vector of variables using ADMM for IEEE 14 bus system

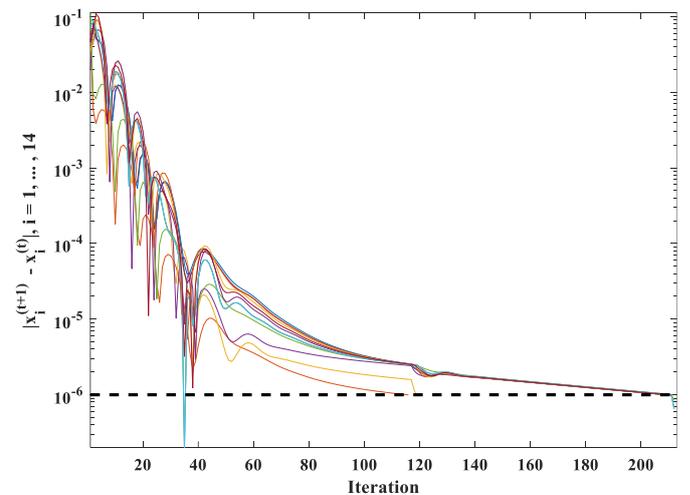

Fig. 3. Convergence curve for vector of variables using ADMM for IEEE 14 bus system, considering stopping criterion

$\epsilon_1$ is the sum of absolute values of difference between centralized and distributed solution (i.e. $\Sigma |x_{cent} - x_{dist}|$), and $\epsilon_2$ is $max(|x_{cent} - x_{dist}|)$. Computation burden means the time that has been spent by computer to solve the problem in a distributed manner. As stated in (Glavic and Van Cutsem 2013), time delay for each iteration can be considered between 0.1 to 0.5 second. So, we selected data transmission delay of $t_{delay}$ = 0.5 as the worst case. In other words, the overall time can be calculated using the following equation:

$$OT = (t_{delay} \times Iter) + CB. \quad (11)$$

Finally, the objective function value for optimal state variables, which was obtained applying different methods, was evaluated using (1).

*4.2 4 Bus system*

Table 1 indicates numerical results for 4 bus system analysis. The centralized objective value for 4 bus system is 8.4125e-27. Considering the results provided in the Table 1, effect of applying stopping criterion is not so clear, except for Gossip based method, which the overall time shows how the stopping criterion helps to decrease the needed time.

Table 1. Numerical results of 4 Bus system

| Methods | | Iter | $\epsilon_1$ | $\epsilon_2$ | CB | OT | OV |
|---|---|---|---|---|---|---|---|
| Matrix splitting | WOSC | 110 | 2.39e-5 | 1.01e-5 | 55.16 | 5.66 | 4.1e-5 |
| | WSC | 109 | 3.66e-5 | 1.35e-5 | 54.72 | 5.66 | 1.08e-4 |
| Gossip based | WOSC | 467 | 1.11e-6 | 4.92e-7 | 0.165 | 233.67 | 1.45e-6 |
| | WSC | 109 | 3.66e-5 | 1.35e-5 | 0.218 | 54.72 | 1.08e-4 |
| Decomposition | WOSC | 16 | 3.43e-7 | 1.34e-7 | 0.77 | 8.77 | 8.77e-9 |
| | WSC | 16 | 3.43e-7 | 1.34e-7 | 0.77 | 8.77 | 8.77e-9 |
| ADMM | WOSC | 70 | 2.58e-5 | 8.75e-6 | 0.06 | 35.6 | 7.67e-5 |
| | WSC | 70 | 2.58e-5 | 8.75e-6 | 0.06 | 35.6 | 7.67e-5 |

*4.1 14 Bus system*

Table 2 provides the detailed numerical results for IEEE 14 bus system. It is to be noted that system data and area specification for IEEE 14 bus system is adapted from (Kekatos et al. 2017). The DC centralized state estimation objective value for IEEE 14 bus system is *10.0524*.

Table 2. Numerical results of IEEE 14 Bus system

| Methods | | Iter | $\epsilon_1$ | $\epsilon_2$ | CB | OT | OV |
|---|---|---|---|---|---|---|---|
| Matrix splitting | WOSC | 1024 | 1.26e-3 | 1.31e-4 | 8.4546 | 529.45 | 10.0565 |
| | WSC | 927 | 4.6e-3 | 4.16e-4 | 8.6483 | 472.15 | 10.1307 |
| Gossip based | WOSC | 2217 | 2.59e-3 | 3.15e-4 | 0.56 | 1109.06 | 10.0689 |
| | WSC | 1870 | 1.27e-2 | 1.22e-3 | 0.87 | 935.87 | 10.5662 |
| Decomposition | WOSC | 40 | 5.61e-4 | 5.61e-4 | 2.83 | 22.83 | 10.5615 |
| | WSC | 40 | 5.61e-4 | 5.61e-4 | 2.87 | 22.87 | 10.5615 |
| ADMM | WOSC | 245 | 2.43e-3 | 2.43e-3 | 0.4282 | 122.93 | 12.036 |
| | WSC | 213 | 2.3e-2 | 2.97e-3 | 0.3829 | 106.88 | 11.8174 |

*4.3 118 Bus*

Table 3 presents numerical results for IEEE 118 bus system. It should be mentioned that, the topology of distributed IEEE 118 bus system is adopted from (Xia et al. 2019). The objective value for IEEE 118 bus system is *102.7758*. In contrast to the 4 bus system, in bulk power systems, like IEEE 118 bus, the effects of considering stopping criterion are more visible.

Table 3. Numerical results of 118 Bus system

| Methods | | Iter | $\epsilon_1$ | $\epsilon_2$ | CB | OT | OV |
|---|---|---|---|---|---|---|---|
| Matrix splitting | WOSC | 63301 | 2.95 | 6.3e-2 | 7319.35 | 39969 | 396.28 |
| | WSC | 39744 | 7.19 | 7.6e-2 | 391074 | 23782 | 2194.4 |
| Gossip based | WOSC | 58811 | 17.01 | 0.3 | 98.5 | 29504 | 12130 |
| | WSC | 37163 | 17.8 | 0.33 | 100.04 | 18681 | 42593 |
| Decomposition | WOSC | 190 | 5.42e-2 | 2.71e-3 | 149.85 | 244.8 | 109.69 |
| | WSC | 190 | 5.42e-2 | 2.71e-3 | 149.85 | 244.8 | 109.69 |
| ADMM | WOSC | 1621 | 1.11 | 1.6e-2 | 5.875 | 816.3 | 142.57 |
| | WSC | 998 | 1.01 | 1.4e-2 | 3.97 | 502.9 | 137.01 |

An issue that is essential to be considered is the amount of data needed to be transmitted for each method. Both matrix splitting and gossip based methods need 2-hop (or neighbour of neighbour; check (Minot and Li (2015) for detailed definition) when there is a power injection in the neighbouring node and this may cause requesting more data from the neighbouring areas, which is against the privacy and security considerations. While the other two methods do not require such a data. Another issue in decomposition method for bulk systems (like 118 bus system); there might be some single steps, the solver traps into local minima, which makes this method inappropriate for realistic applications. There is another point that sometimes CB in WSC is a little more than WOSC. In addition to slight error of computer in reporting process time, it may happen due to extra criteria we bring in to the simulation coding (which is negligible). Another matter needs to be considered is delay time, which plays an important role in specifying the weight of iteration number. If we increase it, number of iteration will have a vital role in overall consumed time, but if we decrease it (to milliseconds), number of iterations can be negligible as well. This shows the important role of communication infrastructure in distributed methods implementation.

Considering obtained numerical results and demonstrated figures, it is obvious that applying this stopping criterion will speed up the process. The benefit of stopping criterion gets more obvious when the size of the system increases. For example in ADMM method, for 4 bus system, the iterations number are the same. Considering IEEE 14 bus system, around 15% improvement in iteration number is obtained, while this amount increases to 38% in case of IEEE 118 bus system.

Although matrix splitting and gossip based methods might have good objective value but considering iteration number and data privacy they are not a good choice. In case of less information usage and considering consumed time, decomposition and ADMM methods are good choice even by increase in the size of the system, we can see that ADMM or decomposition methods are accurate as well. Now that the choice has been narrowed down to ADMM and decomposition, considering Table 3, one may select decomposition method as the best one, but this method needs a solver, which there is not guarantee to reach to the best solution due to the problems explained above. As shown in Fig. 1, the phase angles obtain with and without stopping criterion are quite the same, which shows the effectiveness of the stopping criterion.

All in all, making a final decision about the best method is not an easy task and is highly dependent on the need of method applicant (power utility), but base on the available results in this research, ADMM with help of stopping criterion has the most accepted performance.

## 6. CONCLUSIONS

In present work, the application of stopping criterion was development and addressed for DPSSE. We selected four recent and well-known methods in this area and analysed their application on 4, 14 and 118 bus systems. The results of the paper are threefold:

- A distributed stopping criterion for termination of local optimizer and data transmission was proposed.

- The implementation of method was done in combination with recent and well-known distributed optimization approaches for DPSSE problems. The typical test cases (4, 14 and 118 Bus systems) were studied with introduced performance indicators.

- Based on the detailed discussion presented in Section V, the examples demonstrate the increase of effectiveness of the designed stopping criterion with the increase of complexity of the grid.

Considering the obtained results, ADMM based DPSSE improved by stopping criterion has shown the most promising performance compared to the other presented methods. The development of combination of fast DPSSE technique with more advanced stopping criterion, adaptive to the problem complexity, is a direction for future research.